\documentclass[twocolumn,prb,aps,superscriptaddress,showpacs]{revtex4-1}

\usepackage{amsmath,amssymb}
\usepackage{graphicx}
\usepackage{xcolor}
\usepackage{ulem}

\begin{document}

\title{\textbf{Lifshitz Transition and Metamagnetism: Thermoelectric Studies of CeRu$_{2}$Si$_{2}$}}

\author{M. Boukahil}
\affiliation{Univ. Grenoble Alpes, INAC-SPSMS, F-38000 Grenoble, France}
\affiliation{CEA, INAC-SPSMS, F-38000 Grenoble, France}
\author{A. Pourret}
\email[E-mail me at: ]{alexandre.pourret@cea.fr}
\affiliation{Univ. Grenoble Alpes, INAC-SPSMS, F-38000 Grenoble, France}
\affiliation{CEA, INAC-SPSMS, F-38000 Grenoble, France}
\author{G. Knebel}
\affiliation{Univ. Grenoble Alpes, INAC-SPSMS, F-38000 Grenoble, France}
\affiliation{CEA, INAC-SPSMS, F-38000 Grenoble, France}
\author{D. Aoki }
\affiliation{Univ. Grenoble Alpes, INAC-SPSMS, F-38000 Grenoble, France}
\affiliation{CEA, INAC-SPSMS, F-38000 Grenoble, France}
\affiliation{Institute for Materials Research, Tohoku University, Oarai, Ibaraki, 311-1313, Japan}
\author{Y. \=Onuki}
\affiliation{Faculty of science, University of the Ryukyus, Nishihara, Okinawa 903-0213, Japan}
\author{J. Flouquet}
\affiliation{Univ. Grenoble Alpes, INAC-SPSMS, F-38000 Grenoble, France}
\affiliation{CEA, INAC-SPSMS, F-38000 Grenoble, France}

\date{\today }

\begin{abstract}
We report field and temperature dependent measurements of the thermoelectric power (TEP) across the pseudo-metamagnetic transition (MMT) in CeRu$_{2}$Si$_{2}$. We applied the thermoelectric gradient parallel and perpendicular to the field along the $c$ axis of the tetragonal crystal which is the easy magnetization axis. At the MMT at $H_m$=7.8~T, a strong anomaly in the TEP is observed for both configurations with opposite signs. The anomaly at $H_m$ becomes a cascade of anomalies at very low temperature which seems to be a generic  feature of the TEP at a change in the topology of the Fermi surface (FS) in heavy Fermion multiband systems. Simultaneously, quantum oscillations in the magnetic field dependence of the TEP are observed for both configurations below and above the MMT.
\end{abstract}


\maketitle

During two decades, magnetic quantum phase transitions (QPT) from antiferromagnetic (AF) or ferromagnetic (FM) phases to paramagnetic (PM) ground states have been discussed mainly in the Doniach frame\cite{Doniach1977} or in the picture of the itinerant spin fluctuations\cite{Moriya1985}. These approaches neglect the possibility of a Fermi surface (FS) reconstruction at the QPT. However, in some heavy-fermion compounds a FS reconstruction has been shown directly by quantum oscillations experiments. 
Prominent examples are the AF compounds CeRh$_{2}$Si$_{2}$\cite{Araki2002} and CeRhIn$_{5}$\cite{Shishido2005} where the evolution of the FS has been studied as a function of pressure through their magnetic quantum critical points. The rapid change of the Hall coefficient in the heavy-fermion compound YbRh$_2$Si$_2$ as a function of field through the critical field $H_c$, where the antiferromagnetic order is suppressed, has been also interpreted as a signature of FS reconstruction\cite{Paschen2004, Friedemann2010}. These observations led to the development of unconventional models such as the breakdown of Kondo effect at the QPT\cite{Si}. 
The emerging picture is that a variation from small FS to a large FS through the critical pressure $P_{c}$ or the critical magnetic field $H_{c}$ in AF systems can be observed near the magnetic quantum criticality\cite{Gegenwart2008}.

However, an unambiguous proof of a FS change and furthermore its complete determination by quantum oscillation experiments or ARPES is often very difficult. Thus, a confirmation requires a convergence of various macroscopic and microscopic measurements. Among them the thermoelectric power (TEP) is a very powerful probe as it is directly linked to the energy ($\epsilon$) derivative of the electrical conductivity $\sigma (\epsilon)$ at low temperature\cite{Barnard1972}:
\begin{equation}
S=-\frac{\pi^{2}}{3}\frac{k_{B}^{2}T}{e} \left(\frac{\partial \ln\sigma(\varepsilon)}{\partial \epsilon}\right) _{\varepsilon_{F}}
\end{equation} 
Thus, the TEP is directly related to the derivative of the density of states $N(\epsilon)$. The strength of the TEP to detect FS singularities has been demonstrated clearly three decades ago on simple metals notably in the study of Lifshitz transitions\cite{Blantera1994} which are topological transitions of the FS. They do not break any symmetry and appear as a crossover at finite temperature, but will be  quantum phase transitions at $T=0$. In multiband systems like heavy fermion compounds the TEP response is complex. The thermoelectric response is the sum of the contribution of every subband weighted by its relative conductivity\cite{Miyake2005}. But the signature of electronic instabilities can be followed continuously with clear anomalies.

The aim of this article is to present a complete study of the TEP in  the Ising-type heavy fermion compound CeRu$_2$Si$_2$ where a FS change under magnetic field is already well established by de Haas-van Alphen (dHvA) experiments\cite{Aoki1993, Takashita1996, Julian, Aoki2014}.

CeRu$_2$Si$_2$ crystallizes in the tetragonal ThCr$_2$Si$_2$-type structure. The $c$ axis is the easy magnetization axis.  It is a PM heavy fermion compound located just on the PM border of the AF quantum critical point ($P_{c}\sim$ -0.3~GPa)\cite{Flouquet2005a}. Doping with La or Ge has a negative pressure effect and stabilizes the AF order. On approaching the critical pressure from the AF side, the metamagnetic critical field suppressing the AF order terminates at a quantum critical end point $H^{*}_{c}\sim 4~T$ \cite{Flouquet2005a,Fisher1991, Paulsen1990}. The closeness of the pure PM compound CeRu$_2$Si$_2$ to the magnetic instability induced that for a field closely related to $H^{*}_{c}$ a sharp continuous pseudometamagnetic cross over is observed at $H_{m}\sim 7.8 T$ at ambient pressure. The MMT is associated to a strong enhancement of the Sommerfeld coefficient at $H_{m}$ \cite{Flouquet2005a,Fisher1991, Paulsen1990, Aoki1998}.

 We present TEP measurements on CeRu$_2$Si$_2$ for both configurations, heat current transverse ($J_{Q} \!\parallel \! a, H\! \parallel\! c$) and longitudinal to the applied magnetic field ($J_{Q}\! \parallel \!c, H\! \parallel\! c$) for temperatures down to 120~mK and magnetic fields up to 16~T.  In addition,  resistivity measurements have been performed for both configurations down to 30~mK and magnetic fields up to 13~T on the same single crystals. The TEP shows a strong anomaly at $H_m$ but with different signs depending on the heat current direction with respect to the magnetic field. The strong anisotropy of the TEP at low temperature is coupled to the magnetoresistivity. The high quality of the crystal allows to observe the quantum oscillations of the "light" quasiparticles and detect their changes through $H_{m}$.\\
 
 High-quality single crystals of CeRu$_2$Si$_2$ were grown using the Czochralski pulling method in a tetra-arc furnace. Two different samples have been used: sample 1 had a residual resistance ratio (RRR) of 160 and sample 2 a RRR of 100. The magnetic field is applied along the easy magnetization $c$ axis. Heat and charge currents are applied along [100] in the basal plane for sample 1 (transverse configuration) and along [001] for sample 2 (longitudinal configuration). TEP experiment has been performed by using a "one heater, two thermometers" set-up in a dilution refrigerator down to 120~mK and up to 16~T. Thermometers and heater are thermally decoupled from the sample holder by highly resistive manganin wires. The temperature and field dependence of the TEP has been measured by averaging the TEP voltage during several minutes depending of the temperature with and without thermal gradient. To observe quantum oscillations in the TEP, the field has been swept continuously and the TEP measurements have been obtained by applying a constant power to the heater in order to obtain the thermal gradient during the field sweep. The thermoelectric voltage obtained for zero thermal gradient is taken at the beginning and at the end of the sweep. Resistivity measurements have been performed on the same crystals down to 30~mK and fields up to 13~T by a four point lock-in technique using a low temperature transformer to improve the signal to noise ratio.

\begin{figure}[h!]
\begin{center}
	\includegraphics[width=8cm]{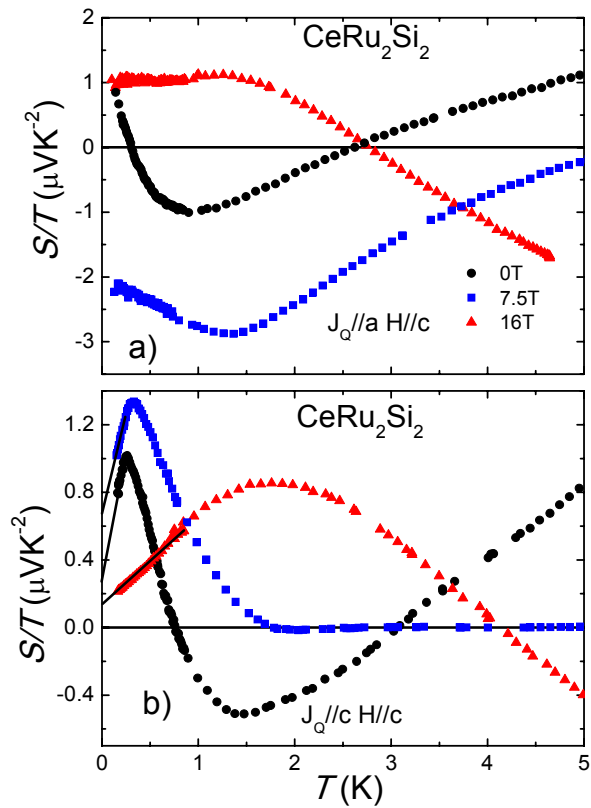}
	\caption{\label{STvsT} (Color online). Temperature dependence of the TEP divided by the temperature at different magnetic fields, for transverse a) and longitudinal b) thermal flow configurations. The presence of a peak in $S/T$ at 260~mK for the longitudinal underlines the difficulty of extrapolating correctly $S/T$ when $T\rightarrow 0$ (black line).}
\end{center}
\end{figure}

 \begin{figure}[h!]
	 \begin{center}
	\includegraphics[width=8cm]{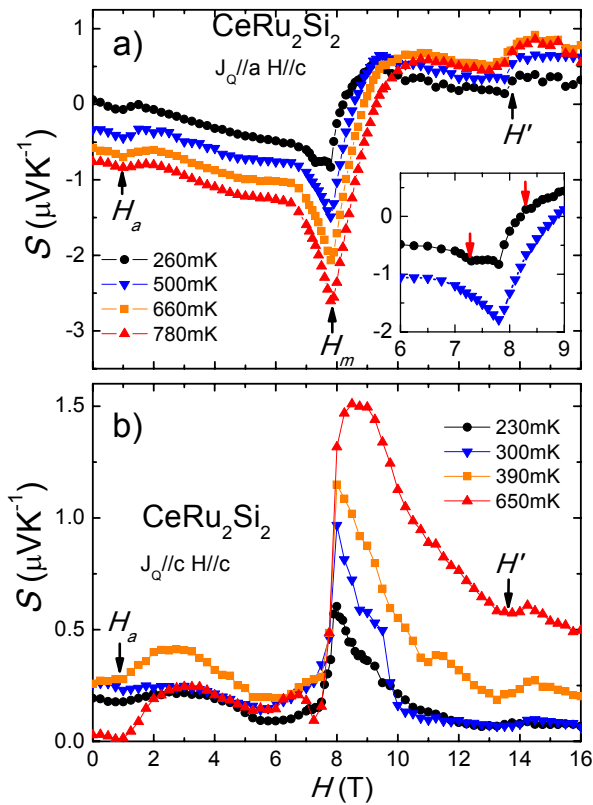}
	
	\caption{\label{SVsB} (Color online). Magnetic field dependence of the TEP at different temperatures for transverse a) and longitudinal b) configuration. $S$ shows a clear negative (positive) peak at the pseudo-metamagnetic field $H_{m}=7.8$~T for the transverse (longitudinal) configuration. The anomaly around 1~T was reported and is attributed to the presence of spin fluctuations \cite{Mignot1991}. The small jump of $S$ at 13.5~T seems to correspond to a softening of the longitudinal mode observed by ultrasound experiment \cite{Yanagisawa2002}. Inset of a): At 260~mK, TEP shows furthers anomalies inside the MMT as pointed by the two red vertical arrows.}
	\end{center}
\end{figure}

The temperature variation of the TEP divided by temperature $S/T $ at different magnetic fields  is represented for the thermal heat current $J_{Q} \parallel a$  and $J_{Q} \parallel c$ in Fig.~\ref{STvsT}. For both configurations, the TEP shows a complex temperature dependence with different extrema and sign changes. On cooling for the transverse heat current, $S(T)/T$  changes sign from positive to negative at $T \sim 2.7$~K, has a minimum at $T\sim 1$~K, gets positive again below $T\sim 0.3$~K and increases down to the lowest temperature of the experiment (100~mK). In difference, for a heat current parallel to the applied field $S (T)/T $ changes sign from positive to negative already at 3~K, has a minimum at 1.4~K, is positive below 750~mK and shows another sharp maximum at $T \sim 260$~mK. Such differences in the temperature dependence of  $S (T)/T$ with respect to the direction of the heat current have been already reported in the first study on single crystals \cite{Amato1988}. However, in that study (RRR$<50$) for a transverse heat current, $S (T)$ is always positive, indicating that the TEP is very sensible to the sample quality.
The positive sign of of the TEP at very low temperature for both configurations at $H=0$  is in agreement with the observation that the TEP in Ce based heavy-fermion compounds is often positive in the limit $T\rightarrow 0$~K. In a spherical single band picture, the TEP goes linearly with $T$ and the ratio $q=\frac{SN_{Av}e}{T \gamma}$, where $N_{Av}$ is the Avogadro's number and $\gamma$ is the Sommerfeld coefficient, is directly related to the inverse of the number of heat carriers per formula unit \cite{Behnia2004a}. However, CeRu$_2$Si$_2$ is a multiband system and such a simple estimation of the number of charge carriers from the TEP is not possible. Clearly the multiband structure leads to an anisotropy of $S/T$ and also to a different extrapolation of $S/T$ with respect to the heat current in the limit $T \to 0$. 
Furthermore, the presence of a peak in $S/T$ for the longitudinal configuration at $T$=260~mK reveals the necessity of very low temperatures to estimate correctly $S/T$ for $T\rightarrow 0$. The extremely low value of the extrapolated $S/T$ for the longitudinal heat current compared to the specific heat leads to $q$=0.1 corresponding to a large number of carriers in agreement with the small Hall effect \cite{Kambe1996}. Interestingly, the Fermi liquid regime ($S/T =$const) is never achieved in the TEP in difference to the resistivity \cite{Daou2006} (see below) or the specific heat.\cite{Aoki1998}

The temperature dependence of $S/T$ varies strongly under magnetic field  as shown in Fig.~\ref{STvsT}, in particular the temperature dependencies below and above $H_m$ are very different, e.g. for $J_Q \parallel a$ the TEP $S (T)$ at $H=16$~T is negative for $T > 3$~K  and gets positive when the low field response is negative. Even at the highest field (16~T) the TEP is strongly anisotropic: while $S (T)/T \approx 1$~$\mu$VK$^{-2}$ for the transverse heat current is constant below $T\sim 1.5$~K indicating the Fermi liquid state, $S(T)/T$ for the longitudinal configuration decreases linearly with decreasing temperature below 1.5~K and $S/T \sim 0.15$~$\mu$VK$^{-2}$ for $T\to 0$. 

 \begin{figure}[h!]
	 \begin{center}
	\includegraphics[width=8cm]{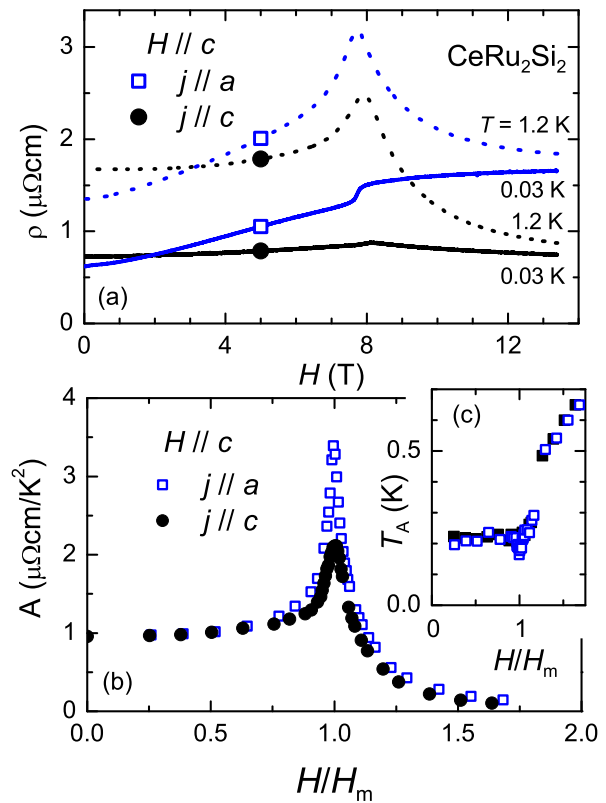}
	\caption{\label{RVsB} (Color online). (a) Magnetic field dependence of the variation of the magnetoresistivity $\rho(H)$ for transverse (squares) and longitudinal (dot) configuration at 30~mK (solid lines) and 1.2~K (dashed lines). (b) The $A$ coefficient of the $T^2$ term of resistivity normalized for $H = 0$ for $j\parallel a$ and $j\parallel c$ as function of field normalized to $H_m$. At $H_m$ the A coefficient shows a sharp peak as function of field. (c) The relative field dependence of the maximum temperature of the $T^2$ dependence of the resistivity for electrical current parallel and perpendicular to the field $H\parallel c$.}
\end{center}
\end{figure}

To clarify the contrasting behavior in the field dependence of the TEP response at different temperature, we show in Fig.~\ref{SVsB} the field dependence for the two configurations at various temperatures. A sharp negative minimum of $S$ as $H\rightarrow H_{m}$ for the transverse configuration by contrast to the positive maximum for the longitudinal configuration. The field dependence of the transverse TEP is in good agreement with previous results \cite{Pfau2012}.
At low field, in both transverse and longitudinal configurations, a small jump occurs at $H_{a} \approx 1$~T. At this field in the La or Ge doped antiferromagnetically ordered systems of the CeRu$_2$Si$_2$ series, a change of the magnetic ordering vector is observed \cite{Mignot1991, Mignot1991b}. Thus we suspect that the relative weight between the three AF hot spots observed in the pure compound \cite{Kadowaki} changes at $H_{a}$. 
For both configurations, a strong variation of the TEP occurs through $H_m$. Furthermore, clearly above $H_m$ another jump in $S$ is observed for $H^\prime\sim 13.5$~T which coincides with a  tiny softening of the elastic constants detected previously in ultrasonic experiments \cite{Yanagisawa2002}. From  $S(H)$ at different temperatures, a cross boundary between nearly AF phase ($H \leq H_{m}$)) and the polarized paramagnetic (PPM) phase ($H \geq H_{m}$) can be drawn in rather good agreement with that derived from previous thermal expansion measurements \cite{Paulsen1990}.

In the inset of Fig.~\ref{SVsB}, $S(H)$ at very low temperatures is represented near $H_m$. By reducing the temperature, the TEP the transition significantly sharpens and at lowest temperature ($T \leq300$~mK) shows several kinks inside the transition indicated by red vertical arrows suggesting a cascade of Lifshitz transitions. Similar behavior of the TEP has been observed in YbRh$_{2}$Si$_{2}$ at the well established Lifshitz transition ($H_{0}$=9.5~T) \cite{Pourret2013b, Pfau2013, Naren2013} where at least 7 anomalies in the TEP have been reported \cite{Pourret2013b} and direct evidence of magnetic field change of FS in  YbRh$_{2}$Si$_{2}$  was proved \cite{Rourke2008a}. Such a complex behavior of the TEP close to a Lifshitz-like anomaly with additional fine structures appears as a consequence of the multiband character of heavy fermion FS.

The huge difference between the field dependence of the TEP for the two configurations is clearly associated with the magnetoresistivity response. 
Figure \ref{RVsB}a) shows the field variation of the resistivity at $T=30$~mK and $T=1.2$~K for electrical current parallel and perpendicular to the magnetic field.
The magnetoresistivity in the transverse configuration is in excellent agreement with previously published data \cite{Daou2006}. At $H_m$ the magnetoresistivity shows a strong maximum at 1.2~K which sharpens when lowering temperature. Below 300~mK the peak vanishes and turns into a step-like increase as the magnetoresistivity changes on entering in the quantum limit where the magnetoresistivity is dominated by the FS topology. In difference, for the longitudinal magnetoresistivity the peak at $H_m$ is preserved down to the lowest temperature of 30~mK. As expected, the longitudinal magnetoresistivity is rather weak while the transverse magnetoresistance increases by more than a factor of two up to 13~T, mainly due to the jump at $H_m$ and large orbital effects. Figure \ref{RVsB}b) indicates the normalized field dependence of the $A$ coefficient of the resistivity, $\rho = \rho_0 + AT^2$. A peak of $A$ at $H_m$ occurs for both configurations;  however for $j \parallel a$ the coefficent $A$ is almost two times higher than for  $j \parallel c$. It may occur that the stronger increase of $A(H)$ on approaching $H_{m}$ for $j \parallel a$ compare to $j \parallel c$ may be the result of the realization of the quantum limit for many orbits for $j \parallel a$ and thus the usual hypothesis of a regime dominated by collisions is invalid. An open possibility is that the strong Ising magnetic anisotropy of Ce atoms is associated with different spatial hybridization. In Fig.~\ref{RVsB} (c) we have also drawn the relative field dependence of $T_A$, the temperature below which a $T^2$ dependence  is observed. In the field range below $H_m$, $T_A \approx 200$~mK is almost field independent. At the MMT, $T_A$ has a very sharp minimum, but does not collapse which clearly marks that the pseudo-metamagnetism does not reach to the AF magnetic quantum criticality. Above $H_m$ the $T^2$ range increases strongly and extends to almost 700~mK for 13~T, independent of the current direction. 

Obviously, the difference in the electronic scattering has great consequences on the total TEP response taking into account the different magnetic field response of the respective electrical conductivity of each band. For a thermal gradient along the $c$ axis, the TEP response remains always positive and in a simple band picture it appears dominated by its hole response in good agreement with dHvA results \cite{Aoki1993,Takashita1996}. Indeed, independent of the field,  the heaviest carriers observed in the dHvA experiments belongs to a hole-like FS.  Following this simplified approach in the transverse configuration, the TEP through $H_m$ seems to come from electron. De facto, FS reconstructions of the "light" electron orbit are also detected \cite{Aoki1993, Takashita1996}. This simple picture, often used in the literature, is based on strong approximations. Indeed in multiband systems, the sign of the TEP is proportional to the derivative of the density of state of each band with respect to the energy at the Fermi energy and there is no simple correspondence between the sign of the TEP and the sign of the heat carriers.

 \begin{figure}[h!]
 \begin{center}
	\includegraphics[width=8cm]{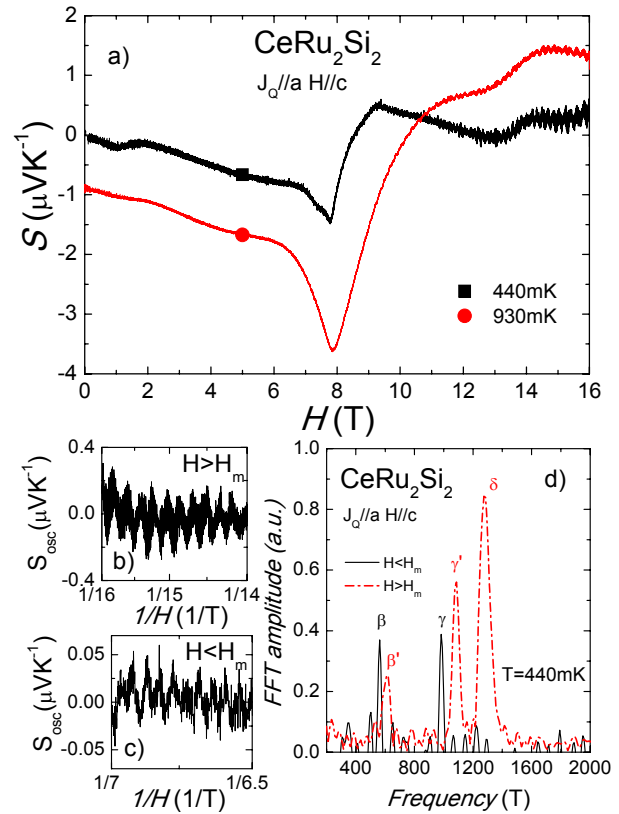}
	\caption{\label{SosVsB} (Color online). a) Field dependence of the TEP at 440~mK (black) and 930~mK (red). The TEP presents quantum oscillations below and above the MMT. b-c) Oscillatory part of the TEP after subtraction of a polynomial background as function of $1/H$ above $H_m$ for the field range from 14~T to 16~T  and below $H_m$ for the range from 6.5~T to 7~T. d) Frequencies of the oscillating TEP at 440~mK below (black) and above (red) the MMT for the transverse configuration. The different branches $\beta$, $\gamma$ for H$\leq$ H$_{m}$ and $\beta'$, $\gamma'$, $\delta$ for $H$ $\geq$ $H_{m}$ are in good agreement with dHvA measurements \cite{Aoki1993}. }
\end{center}
\end{figure} 
Direct evidence of a FS reconstruction is also found in the TEP as we were able to detect superimposed quantum oscillation in the TEP below and above $H_m$ for both configurations (see Fig.~\ref{SosVsB} for the transverse configuration). 
Only recently quantum oscillations in the TEP had been observed in heavy fermion systems and they are reported only in few metallic materials like CeZn$_{11}$ \cite{Hodovanets2013}. Thermoelectric quantum oscillations were first observed in pure metals\cite{Papastaikoudis1979} and later in semimetallic systems, mostly in the Nernst coefficient, like in Bi \cite{Behnia2007d}  or in Graphite \cite{Zhu2009}.  The presence of oscillations below and above $H_{m}$ allows us to make a precise study of the evolution of the FS through the MMT transition. Figure \ref{SosVsB}a) shows the TEP for a thermal heat current along the $a$ axis as a function of field for $T=440$~mK and $T=930$~mK by continuously sweeping the field. Figure \ref{SosVsB}b) shows the oscillatory part of the  TEP at $T=440$~mK represented as a function of $1/H$ between 14~T and 16~T and in c) in the field range from 6.5~T to 7~T  after subtraction of a polynomial background. Figure \ref{SosVsB}d) displays the Fourier transform spectrum of TEP for $T=400$~mK for $H<H_{m}$ (black full line) and $H>H_{m}$ (red dashed dot). The observed frequencies below and above $H_m$ are in excellent agreement with previous dHvA oscillations experiment \cite{Aoki1993}.  Below $H_{m}$ the observed branches can be explained by band calculation \cite{Yamagami1993, Suzuki2010a} supposing the $4f$ electrons itinerant. The light $\beta$ and $\gamma$ branches are attributed to  ellipsoidal hole surfaces centered at the Z point of the Brillouin zone. We had not been able to detect the heavy branches $\kappa$ and $\psi$, ($m^\star \sim 11$m$_0$ and $\approx 200$m$_0$, respectively) due to the limited temperature range of the TEP measurement. Above  $H_{m}$, the three FS branches have been observed, $\beta'$, $\gamma'$, and $\delta$. 
The large $\omega$ hole orbit, which is commonly observed in isostructural compounds like CeRu$_2$Ge$_2$ or LaRu$_2$Si$_2$ for which the $4f$ electron is localized or absent, could not be detected. Importantly, our TEP measurement give no indication of any abrupt change of the FS due to a first-order like change, but they are in agreement with a continuous change of the FS\cite{Daou2006}.

The present data can be compared with recent results obtained on Ce(Ru$_{0.92}$Rh$_{0.08}$)$_2$Si$_{2}$ crystals where the critical field $H_c$ to suppress AF order is decoupled from $H_{m}$ as the Rh doping modifies the dominant hot spot of the antiferromagnetic correlations. For both configurations ($J_{Q} \parallel a$  and $J_{Q} \parallel c$ ) sharp anomalies of the TEP appear at $H_{c}$ and $H_{m}$. More surprisingly no main anisotropy is detected at $H_{m}$. The shape of $S$ at $H_{m}$ for both configurations is rather similar than that observed for the transverse configuration of the pure CeRu$_{2}$Si$_{2}$ lattice.  Drastic differences in the anisotropy of the TEP at $H_{m}$ between CeRu$_{2}$Si$_{2}$ and Ce(Ru$_{0.92}$Rh$_{0.08}$)$_2$Si$_{2}$ appears related to their contrasting magnetoresistivity \cite{Machida2013}. Of course, for Rh doping the residual resistivity at $H=0$ is already large ($\rho_0 \approx 10 \mu\Omega$cm) and thus no drastic change in the resistivity is induced between the longitudinal and the transverse configuration. In CeRu$_{2}$Si$_{2}$ as well as in YbRh$_{2}$Si$_{2}$, the high magnetic field electronic instability appears when the field induced magnetization reaches a critical value \cite{Flouquet2005a, Pfau2013}. The FS reconstruction is not associated directly with metamagnetism. Indeed MMT occurs in CeRu$_{2}$Si$_{2}$ but not in YbRh$_{2}$Si$_{2}$. The key ingredient is the degree of magnetic polarization of the bands. The features favors a scenario \cite{Miyake2006, Daou2006} where the Zeeman effect on one sheet of the spin split FS shrinks to zero volume leading to a Lifshitz transition. It was argued two decades ago that the 4f electrons itinerant below $H_{m}$ will become localized above $H_{m}$ \cite{Aoki1993,Takashita1996,Julian}. As there is one itinerant electron missing in the localized case, the image is that of small FS (above $H_{m}$) by comparison to the large FS (below $H_{m}$). Restricted experimented evidences were the detection of orbits above $H_{m}$ predicted in band structure calculation \cite{Yamagami1993, Suzuki2010a} assuming the 4f electrons localized. However only few orbits are observed above $H_{m}$ and large parts of the FS are not observed. The observed orbits cannot explain the thermodynamic  \cite{Flouquet2005a, Sakikabara}  and transport\cite{Daou2006, Kambe1996} properties above  $H_{m}$. This gives strong support  for the persistence of the 4f itinerancy through $H_{m}$ in agreement with the occurrence of a Lifshitz transition of polarized band.

To summarize, we present a detailed study of the TEP of CeRu$_2$Si$_2$ for heat current applied parallel or transverse to the magnetic field. These longitudinal and transverse measurements, with respect to the applied magnetic field along the $c$ axis, reveal quite contrasting responses.
Strong anomalies are detected for both configurations at the MMT $H_m$.
 It is clearly associated with a large FS reconstruction which occurs at $H_{m}$. The anomalies inside the transition at $H_{m}$ show that the FS evolution may occur in field window as recently detected for YbRh$_{2}$Si$_{2}$.\cite{Pourret2013, Pfau2013} Furthermore additional anomalies have been detected at $H_a\sim1$~T and $H^\prime\sim13.5$~T. A crude analysis on the magnetoresistivity response is made to understand qualitatively $S(H)$. An interesting theoretical point will be to elucidate the possible role of the anisotropic hybridization on the TEP anisotropy. The combination of improvements in the crystal quality and in reducing the signal to noise level of the measurement lead to observe directly significant changes in the quantum oscillation frequencies of the light FS orbit. For CeRu$_{2}$Si$_{2}$ the FS in the low field paramagnetic phase ($H<H_m$) is excellently known.  Clearly, direct evidence is given for a FS reconstruction through the critical pseudo-metamagnetic field $H_m$ and a broad data basis of various measurements, including the spin dynamics exists. Thus the observed effects in the TEP can serve as reference for future experimental and theoretical studies in heavy-fermion systems and in particular in the detection of FS instabilities. Our result must push to quantitative modeling of CeRu$_{2}$Si$_{2}$. Theoretical discussion on consequences of Lifshitz transition in heavy fermions systems were made for CeIn$_{3}$ on crossing their magnetic bondary \cite{Gorkov} for the mean field study of the heavy fermion metamagnetic transition \cite{Kusminskiy}, for the high field reconstruction of the FS of  YbRh$_{2}$Si$_{2}$ at $H^{*} \sim 12~T$ very far from the critical field $H_{c}=0.066~T$ where the systems switches from AF to PM. Proposal has even been made that the transition at $H_{c}$ may be a Lifshitz one\cite{Hackl}.\\

We thank K.~Miyake and H.~Harima for many useful discussions. This work has been supported by the French ANR (projects PRINCESS), the ERC (starting grant NewHeavyFermion), ICC-IMR and REIMEI.

\bibliographystyle{apsrev4-1}

\end{document}